\documentclass[12pt]{article}
\usepackage{graphicx}

\newcommand\pubdate{\today}

\newcommand\pubnumber{RM3-TH/11-11; CERN-PH-TH/2011-197}

\def\gappeq{\mathrel{\rlap {\raise.5ex\hbox{$>$}} {\lower.5ex\hbox{$\sim$}}}}
\def\lappeq{\mathrel{\rlap{\raise.5ex\hbox{$<$}} {\lower.5ex\hbox{$\sim$}}}}

\def\Title#1{\begin{center} {\Large #1 } \end{center}}
\def\Author#1{\begin{center}{ \sc #1} \end{center}}
\def\Address#1{\begin{center}{ \it #1} \end{center}}

\newcommand\pubblock{\rightline{\begin{tabular}{l} \pubnumber\\
         \pubdate  \end{tabular}}}
\newenvironment{Abstract}{\begin{center}{\bf Abstract}\end{center} \bigskip \begin{quotation}  }{\end{quotation}}
\newenvironment{Presented}{\begin{quotation} \begin{center} 
             PRESENTED AT\end{center}\bigskip 
      \begin{center}\begin{large}}{\end{large}\end{center} \end{quotation}}
\def\Acknowledgements{\bigskip  \bigskip \begin{center} \begin{large}
             \bf ACKNOWLEDGEMENTS \end{large}\end{center}}

\def\beq{\begin{equation}}
\def\eeq#1{\label{#1}\end{equation}}
\def\eeqn{\end{equation}}

\def\beqa{\begin{eqnarray}}
\def\eeqa#1{\label{#1}\end{eqnarray}}
\def\eeqan{\end{eqnarray}}

\let\bar=\overbar

\def\L{{\cal L}}

\def\Dslash{\not{\hbox{\kern-4pt $D$}}}
\def\dslash{\not{\hbox{\kern-2pt $\del$}}}

\def\msb{{\bar{\ssstyle M \kern -1pt S}}}

\begin{document}
\begin{titlepage}
\pubblock

\vfill


\Title{Concluding Talk: A Theorist Overview on Particle Physics}
\vfill
\Author{G. Altarelli}  
\Address{Dipartimento di Fisica, Universita' di Roma Tre\\
Rome, Italy\\
and\\
CERN, Department of Physics, Theory Division \\  
CH-1211 Gen\`eve 23, Switzerland\\
{\rm E-mail: guido.altarelli@cern.ch}}
\vfill


\begin{Abstract}

This is a Concluding Talk, not a Summary of the FPCP 2011 Conference. I will first make some comments on the status and the prospects of particle physics and then review some of the
highlights that particularly impressed me at this Conference (a subjective choice). 

\end{Abstract}

\vfill

\begin{Presented}
The Ninth International Conference on\\
Flavor Physics and CP Violation\\
(FPCP 2011)\\
Maale Hachamisha, Israel,  May 23--27, 2011
\end{Presented}
\vfill

\end{titlepage}
\def\thefootnote{\fnsymbol{footnote}}
\setcounter{footnote}{0}
%


\section{The General Context}

The LHC is working very well: on the day of my talk (May 27th 2011) a total of 0.42 $fb^{-1}$ of integrated luminosity had been collected. While writing this Proceedings (August 16th 2011) the score is up to 2.5 $fb^{-1}$. This is great news because particle physics is in a deadlock
and needs to be restarted. One eagerly waits for the answers from the LHC to a number of big questions. As well known, the main physics issues at the LHC, addressed by the ATLAS and CMS collaborations, are: 1) the experimental clarification of the Higgs sector of the electroweak (EW) theory, 2) the search for new physics at the weak scale that, on conceptual grounds, one predicts should be in the LHC discovery range, and, hopefully, 3) the identification of the particle(s) that make the Dark Matter in the Universe, in particular if those are WIMPs (Weakly Interacting Massive Particles). In addition the LHCb detector will be devoted to the study of precision B physics, with the aim of going deeper into the physics of the Cabibbo-Kobayashi-Maskawa (CKM) matrix and of CP violation. The LHC will also devote a number of runs to accelerate heavy ions and the ALICE collaboration will study their collisions for an experimental exploration of the QCD phase diagram.

The Standard Model (SM) is a low energy effective theory (nobody can believe it is the ultimate theory).
It happens to be renormalizable, hence highly predictive and is extremely well supported by the data.
However, we expect corrections from higher energies, not only from the Grand Unification (GUT) or Planck scales
but also from the TeV scale (LHC!). In fact, even just as a low energy effective theory,
the SM is not satisfactory: while the QCD and the gauge part of the EW theory are fine,
the Higgs sector is so far only a mere conjecture. The experimental verification of the SM cannot be considered complete until the predicted physics of the  Higgs sector is not established by experiment \cite{djou1}. Indeed the Higgs problem is really central in particle physics today \cite{wells}. In fact, the Higgs sector is directly related to most of the major open problems of particle physics, like the flavour problem and the hierarchy problem, the latter strongly suggesting the need for new physics near the weak scale, which could also clarify the Dark Matter identity. 

It is clear that the fact that some sort of Higgs mechanism is at work has already been established. The W's or the Z's with longitudinal polarization that are observed are not present in an unbroken gauge theory (massless spin-1 particles, like the photon, are transversely polarized).The observed W and Z longitudinal degrees of freedom are borrowed from the Higgs sector and are an evidence for it. Also, the couplings of
the weak gauge bosons W$^{\pm}$ and Z to quarks and leptons are indeed experimentally found to be precisely those
prescribed by the gauge symmetry.  To a lesser
accuracy the triple gauge vertices $\gamma$WW and ZWW have also
been found in agreement with the specific predictions of the
$SU(2)\bigotimes U(1)$ gauge theory. This means that it has been
verified that the gauge symmetry is unbroken in the interaction vertices: all currents and charges are indeed symmetric. Yet there is obvious
evidence that the symmetry is instead badly broken in the
masses.  Not only the W and the Z have large masses, but the large splitting of, for example,  the t-b doublet shows that even the global weak SU(2) is not at all respected in the spectrum. Symmetric couplings and totally non symmetric spectrum is a clear signal of spontaneous
symmetry breaking and its implementation in a gauge theory is via the Higgs mechanism. The big remaining questions are about the nature and the properties of the Higgs particle(s): one doublet, more doublets, additional singlets? SM Higgs or SUSY Higgses? Fundamental or composite (if so, of what: of fermions, of WW....)? Pseudo-Goldstone boson of an enlarged symmetry? A manifestation of extra dimensions (fifth component
of a gauge boson, an effect of orbifolding or of boundary
conditions....)? Some combination of the above?
The LHC has been designed to solve the Higgs problem. 

A crucial argument indicating that the solution of the Higgs problem cannot be too far away and very likely in the LHC discovery range is the fact that, in the absence of a Higgs particle or of an alternative mechanism, violations of unitarity appear in scattering amplitudes involving longitudinal gauge bosons (those most directly related to the Higgs sector) at energies in the few TeV range \cite{ref:unit}. A very important issue for the LHC is to identify the mechanism that avoids the unitarity violation: is it one or more Higgs bosons or some new vector boson (like additional gauge bosons WÕ, ZÕ or Kaluza-Klein recurrences or resonances from a strong sector)?

It is well known \cite{mhlow} that in the SM with only one Higgs doublet a lower limit on
$m_H$ can be derived from the requirement of vacuum stability (i.e. that the quartic Higgs coupling $\lambda$ does not turn negative in its running up to a large scale $\Lambda$) or, in milder form, of a moderate instability, compatible with the lifetime of the Universe  \cite{ref:isid}. The Higgs mass enters because it fixes the initial value of the quartic Higgs coupling $\lambda$. For the experimental value of $m_t$ the lower limit is below the direct experimental bound for $\Lambda \sim $ a few TeV and is $M_H> 130$ GeV for $\Lambda \sim M_{Pl}$. Similarly an upper bound on $m_H$ (with mild dependence
on $m_t$) is obtained, as described in \cite{hri}, from the requirement that for $\lambda$ no Landau pole appears up to the scale $\Lambda$, or in simpler terms, that the perturbative description of the theory remains valid up to  $\Lambda$. The upper limit on the SM Higgs mass is clearly important for assessing the chances of success of the LHC as an accelerator designed to solve the Higgs problem. Even if $\Lambda$ is as small as ~a few TeV the limit is $m_H < 600-800~$GeV and becomes $m_H < 180~$GeV for $\Lambda \sim M_{Pl}$. 

Is it possible that the solution of the Higgs problem is not found at the LHC? Here by Higgs we mean the EW symmetry breaking mechanism. Looks pretty unlikely. In fact, the LHC discovery range is large enough: at $14$ TeV it extends up to $m_H < \sim$ 1 TeV, so that the Higgs should be really heavy to escape discovery at the LHC. Radiative corrections indicate a light Higgs (whatever its nature). In the SM we obtain $m_H \lappeq 185~GeV$ at $95\%$ c.l. \cite{ewgr}.
A heavy Higgs would make perturbation theory to
collapse nearby (as we have seen there would be violations of unitarity for $m_H > \sim$ 2-3 TeV)
Such nearby collapse of perturbation theory is very difficult to reconcile
with EW precision tests plus simulating a light Higgs.
The SM good agreement with the data favours forms of new physics that keep at least some Higgs light.
In conclusion it looks very likely that the LHC can clarify the problem of the EW symmetry breaking mechanism. It has been designed for it!

No signal of new physics has been
found neither in electroweak precision tests nor in flavour physics. Given the success of the SM why are we not satisfied with that theory? Why not just find the Higgs particle,
for completeness, and declare that particle physics is closed? The reason is that there are
both conceptual problems and phenomenological indications for physics beyond the SM. On the conceptual side the most
obvious problems are the fact that quantum gravity is not included in the SM, the proliferation of parameters, the puzzles of family replication and of the flavour hierarchies and the hierarchy problem. Among the main
phenomenological hints for new physics we can list the quest for Grand Unification and coupling constant merging, Dark Matter, neutrino masses (which suggest Majorana neutrinos and lepton number L non conservation), 
baryogenesis and the cosmological vacuum energy (a gigantic naturalness problem). 
The computed evolution with energy
of the effective gauge couplings clearly points towards the unification of the electro-weak and strong forces in GUT's at scales of energy
$M_{GUT}\sim  10^{15}-10^{16}~ GeV$ which are close to the scale of quantum gravity, $M_{Pl}\sim 10^{19}~ GeV$.  One is led to
imagine  a unified theory of all interactions also including gravity (at present superstrings provide the best attempt at such
a theory). Thus GUT's and the realm of quantum gravity set a very distant energy horizon that modern particle theory cannot
ignore. Can the SM without new physics be valid up to such large energies? One can imagine that some problems could be postponed to the more fundamental theory at the Planck mass. For example, the explanation of the three generations of fermions and the understanding of fermion masses and mixing angles can be postponed. But other problems must find their solution at the EW scale. In particular, the structure of the
SM could not naturally explain the relative smallness of the weak scale of mass, set by the Higgs mechanism at $\mu\sim
1/\sqrt{G_F}\sim  250~ GeV$  with $G_F$ being the Fermi coupling constant. This so-called hierarchy problem is due to the instability of the SM with respect to quantum corrections. This is related to
the
presence of fundamental scalar fields in the theory with quadratic mass divergences and no protective extra symmetry at
$\mu=0$. For fermion masses, first, the divergences are logarithmic and, second, they are forbidden by the $SU(2)\bigotimes
U(1)$ gauge symmetry plus the fact that at
$m=0$ an additional symmetry, i.e. chiral  symmetry, is restored. Here, when talking of divergences, we are not
worried of actual infinities. The theory is renormalizable and finite once the dependence on the cut off $\Lambda$ is
absorbed in a redefinition of masses and couplings. Rather the hierarchy problem is one of naturalness. We can look at the
cut off as a parameterization of our ignorance on the new physics that will modify the theory at large energy
scales. Then it is relevant to look at the dependence of physical quantities on the cut off and to demand that no
unexplained enormously accurate cancellations arise. 

The hierarchy problem can be put in less abstract terms: loop corrections to the higgs mass squared are
quadratic in the cut off $\Lambda$: $\delta m_h^2 \sim \Lambda^2$. The most pressing problem is from the top loop that produces the largest coefficient for the quadratic term in the cut-off.
If we demand that the correction does not exceed the light Higgs mass indicated by the precision tests, $\Lambda$ must be
close, $\Lambda\sim o(1~TeV)$ (the "little hierarchy problem"). Similar constraints arise from the quadratic $\Lambda$ dependence of loops with gauge bosons and
scalars, which, however, lead to less pressing bounds. So the hierarchy problem demands new physics to be very close (in
particular the mechanism that quenches the top loop). Actually, this new physics must be rather special, because it must be
very close, yet its effects are not clearly visible - the "LEP Paradox" \cite{ref:BS} - now also accompanied by a similar "flavour paradox" \cite{isid} (see the theory talks by Weiler, Grossman and Hiller and the experimental ones by Onuki, Schune, Sahoo and Mohanti). 

As well known, examples of proposed classes of solutions to the hierarchy problem are:

¥ $\bf{Supersymmetry.}$ In the limit of exact boson-fermion symmetry \cite{ref:Martin} the quadratic divergences of bosons cancel so that
only log divergences remain. However, exact SUSY is clearly unrealistic. For approximate SUSY (with soft breaking terms),
which is the basis for all practical models, $\Lambda$ is replaced by the splitting of SUSY multiplets. In particular, the top loop is quenched by partial cancellation with s-top exchange, so the s-top cannot be too heavy. An important phenomenological indication for SUSY is that coupling unification is quantitatively precise in SUSY GUT's and that proton decay bounds are not in contradiction with the predictions. An interesting exercise is to repeat the fit of precision tests in the Minimal Supersymmetric Standard Model with GUT constraints added (CMSSM), also including the additional data on the muon $(g-2)$, the Dark Matter relic density and on b-decay. The result is that the central value of the lightest Higgs mass $m_h$ goes up (in better harmony with the bound from direct searches). The best fit is for moderately large $tan\beta$ and a relatively light SUSY spectrum \cite{ref:sus}. But the rapidly improving bounds from the LHC push the optimal region of parameter space (within the probably too narrow CMSSM) towards smaller probability and larger fine tuning \cite{struFT}. However less fine tuning is necessary if non minimal models are assumed (for a recent example, see \cite{ryk})

¥ $\bf{Technicolor.}$ The Higgs system is a condensate of new fermions. There is no fundamental scalar Higgs sector, hence no
quadratic divergences associated to the $\mu^2$ mass in the scalar potential. This mechanism needs a very strong binding force,
$\Lambda_{TC}\sim 10^3~\Lambda_{QCD}$. It is  difficult to arrange that such nearby strong force is not showing up in
precision tests. Hence this class of models has been disfavoured by LEP, although some special class of models have been constructed a posteriori, like walking TC, top-color assisted TC etc \cite{ref:L-C} and more recently some extra dimensional models based on the AdS/CFT correspondence \cite{con}. 

¥ $\bf{"Little~Higgs"~models.}$  In "little Higgs" models \cite{schm} the symmetry of the SM is extended to a suitable global group G that also contains some
gauge enlargement of $SU(2)\bigotimes U(1)$, for example $G\supset [SU(2)\bigotimes U(1)]^2\supset SU(2)\bigotimes
U(1)$. The Higgs particle is a pseudo-Goldstone boson of G that can only take mass at the 2-loop level, because two distinct
symmetries must be simultaneously broken for this to happen, which requires the action of two different couplings in
the same diagram. Then in the relation between
$\delta m_h^2$ and
$\Lambda^2$ there are an additional coupling and an additional loop factor that imply a larger separation between the Higgs
mass and the cut-off. Typically, in these models one has one or more Higgs doublets at $m_h\sim~0.2~{\rm TeV}$, and a cut-off at
$\Lambda\sim~10~{\rm TeV}$. The top loop quadratic cut-off dependence is partially canceled, in a natural way guaranteed by the
symmetries of the model, by a new coloured, charge 2/3, vectorlike quark $\chi$ of mass around $1~{\rm TeV}$ (a fermion not a scalar
like the s-top of SUSY models). Certainly these models involve a remarkable level of group theoretic virtuosity. However, in
the simplest versions one is faced with problems with precision tests of the SM \cite{prob}. These problems can be fixed by complicating the model \cite{Ch}: one can introduce a parity symmetry, T-parity, and additional "mirror" fermions.  T-parity interchanges the two $SU(2)\bigotimes
U(1)$ groups: standard gauge bosons are T-even while heavy ones are T-odd. As a consequence no tree level contributions from heavy $WÕ$ and $ZÕ$ appear in processes with external SM particles. 
Therefore all corrections to EW observables only arise at loop level. A good feature of T-parity is that, like for R-parity in the MSSM, the lightest T-odd particle is stable (usually a B') and can be a candidate for Dark Matter (missing energy would here too be a signal) and T-odd particles are produced in pairs (unless T-parity is not broken by anomalies \cite{hill}). Thus the model could work but, in my opinion, the real limit of
this approach is that it only offers a postponement of the main problem by a few TeV, paid by a complete loss of
predictivity at higher energies. In particular all connections to GUT's are lost. Still it is very useful as it offers to experiment a different example of possible new physics and the related signals to look for \cite{sign}.

¥ $\bf{Extra~dimensions.}$ In the original approach \cite{led} the idea was that $M_{Pl}$ appears very large, or equivalently that gravity appears very weak,
because we are fooled by hidden extra dimensions (ED), so that the real gravity scale is reduced down to a much lower scale and that effects of extra dimensions could be detectable at energies of
$o(1~TeV)$ ("large" extra dimensions). This possibility is very exciting in itself and it is really remarkable that it is not directly incompatible with experiment but a realistic model has not emerged \cite{Jo}. At present, the most promising set of ED models are those with "warped" metric, which offer attractive solutions to the hierarchy problem \cite{RS}, \cite{FeAa}  and also to the problem of fermion mass ratios (see, e. g. the talk by Weiler) \cite{warpfla}. The hierarchy suppression $m_W/M_{Pl}$ arises from the warping exponential $e^{-kR\phi}$, with $k\sim M_{Pl}$, for not too large values of the warp factor exponent: $kR\sim 12$ (ED are not "large" in this case). The question of whether these values of $kR$ can be stabilized has been discussed in ref.\cite{GW}. An important direction of development is the study of symmetry breaking by orbifolding and/or boundary conditions. These are models where a larger gauge symmetry (with or without SUSY) holds in the bulk. The symmetry is reduced on the 4-dim. brane, where the physics that we observe is located, as an effect of symmetry breaking induced geometrically by orbifolding or by suitable boundary conditions. In particular SUSY GUT models in ED have been studied where the breaking of the GUT symmetry by orbifolding avoids the introduction of large Higgs representations and also solves the doublet-triplet splitting problem \cite{Kaw}, \cite{edgut}. Also "Higgsless  models" have been tried where it is the SM electroweak gauge symmetry which is broken at the boundaries \cite{Hless}. The violation of unitarity associated with the absence of the Higgs exchange is damped by the Kaluza-Klein recurrences of the gauge bosons. In this case no Higgs should be found at the LHC but other signals, like additional vector bosons, should appear.  The main difficulty is represented by the compatibility with the electro-weak precision tests. 

An interesting class of models that combine the idea of the Higgs as a pseudo-Goldstone boson and warped ED was proposed and studied in ref.s \cite{con} where a kind of composite Higgs in a 5-dim AdS theory appears. This approach can be considered as a new way to look at technicolor \cite{ref:L-C} using the AdS/CFT correspondence. In a RS warped metric framework all SM fields are in the bulk but the Higgs is localised near the TeV brane. The Higgs is a pseudo-Goldstone boson and the electroweak symmetry breaking is triggered by 
top-loop effects. In 4-dim the bulk appears as a strong sector.  The 5-dim theory is weakly coupled so that the Higgs potential and EW observables can be computed.
The Higgs is rather light: $m_H < 185~{\rm GeV}$. Problems with EW precision tests and the $Zb \bar b$ vertex have been fixed in latest versions. The signals at the LHC for this model are 
a light Higgs and new resonances at ~1- 2 TeV

In conclusion, note that apart from Higgsless models (if any?) all theories discussed 
here have a Higgs in LHC range (most of them light).

¥ $\bf{Effective~theories~for~compositeness.}$ In this framework \cite {cont}, \cite{comp}, \cite{comp2} a low energy theory from truncation of some UV completion is described in terms of an elementary sector (the SM particles minus the Higgs), a composite sector (including the Higgs, massive vector bosons and new fermions) and a mixing sector. The Higgs is a pseudo-Goldstone boson of a larger broken gauge group. At low energy, the particle content is identical to the SM one: there exists a 
light and narrow Higgs-like scalar but this particle is a composite from some strong dynamics
and a mass gap separates the Higgs boson from the strong sector as 
a result of the Goldstone nature of the Higgs. The 
effective Lagrangian can be seen as an expansion in $\xi = (v/f)^2$  where $v$ is the Higgs vev and $f$ is the typical scale of the strong sector. The parameter $\xi$ interpolates between the SM limit ($\xi = 0$) and the technicolor limit ($\xi = 1$), 
where a resummation of the full series in $\xi$ is needed.
Non vanishing values of $\xi$ can lead to observable signatures of compositeness at the LHC \cite{comp2}). 

¥ $\bf{The~anthropic~solution.}$ The apparent value of the cosmological constant $\Lambda$ poses a tremendous, unsolved naturalness problem \cite{tu}. Yet the value of $\Lambda$ is close to the Weinberg upper bound for galaxy formation \cite{We}. Possibly our Universe is just one of infinitely many (Multiverse) continuously created from the vacuum by quantum fluctuations. Different physics takes place in different Universes according to the multitude of string theory solutions (~$10^{500}$). Perhaps we live in a very unlikely Universe but the only one that allows our existence \cite{anto},\cite{giu}. I find applying the anthropic principle to the SM hierarchy problem excessive. After all we can find plenty of models that easily reduce the fine tuning from $10^{14}$ to $10^2$: why make our Universe so terribly unlikely? By comparison the case of the cosmological constant is a lot different: the context is not as fully specified as the for the SM (quantum gravity, string cosmology, branes in extra dimensions, wormholes through different Universes....)

Supersymmetry remains the standard way beyond the SM. What is unique to SUSY, beyond leading to a set of consistent and
completely formulated models, as, for example, the MSSM, is that this theory can potentially work up to the GUT energy scale.
In this respect it is the most ambitious model because it describes a computable framework that could be valid all the way
up to the vicinity of the Planck mass. The SUSY models are perfectly compatible with GUT's and are actually quantitatively
supported by coupling unification and compatible with proton decay bounds and also by what we have recently learned on neutrino masses. All other main ideas for going
beyond the SM do not share this synthesis with GUT's. The SUSY way is testable, in particular at the LHC, and the issue
of its relevance at the EW scale will be decided by experiment. It is true that we could have expected the first signals of SUSY already at
LEP2, based on naturalness arguments applied to the most minimal models (like those with simplest universality properties at the GUT scale). The absence of signals has stimulated the development of new ideas like those of extra dimensions
and "little Higgs" models. These ideas are very interesting and provide an important set of targets for the LHC
experiments. Models along these new ideas are not so completely formulated and studied as for SUSY and no well defined and
realistic baseline has so far emerged. But it is well possible that they might represent at least a part of the truth and it
is very important to continue the exploration of new ways beyond the SM. New input from experiment is badly needed, so we all look forward to the results of the LHC.

\section{Highlights from FPCP 2011}

I will start my list of highlights with some new results on heavy quark spectroscopy (talks by Santoro, Choi, Gradl, Bondar and Rosner). Some new bottomonium states were observed by Belle: $h_b$(9898) and $h_b$(10260) are neutral, spin-0, P-wave, $\bar b b$, $J^{PC} = 1^{+-}$. In the charmonium sector the neutral state X(3872), can probably be interpreted as DD* with some admixture of $\bar c c$. More intriguing are the states $Z_b$(10610), $Z_b$(10650), also observed by Belle. Those are charged states containing a $\bar b b$ pair, but they are certainly exotic, in the sense that, being charged, they cannot be pure $\bar b b$ but must be $\bar b b \bar q' q$. Thus they could be interpreted as tetraquarks but also as BB* and B*B* molecular states. It is remarkable that these states were predicted by Karliner and Lipkin in '08 \cite{kar}. The interpretation in terms of tetraquarks or molecular states are often in competition. The lightest nonet of scalar states of around 1 GeV of mass may appear as tetraquark states, because of their inverted spectrum, with strange states lighter than non strange ones (recall that a light di-quark behaves as a strange quark $q^3 \sim \epsilon^{123}q_1q_2$), but are strongly coupled to $\pi \pi$, K$\pi$ and KK.  

On neutrino oscillations, mass and mixing (see the talks by Evans, Zimmermann, Kolomensky and Kayser) the main recent development was the coming back
of sterile neutrinos and the first results from T2K \cite{t2k}. On the evidence for sterile neutrinos a number of hints have been reported in the last months. They do not make yet a clear evidence but certainly pose an experimental problem that needs clarification. First, there is the MiniBooNE experiment that in the antineutrino channel reports an excess of events supporting the LSND oscillation signal (originally observed with antineutrinos). In the neutrino channel MiniBooNE did not observe a signal in the LSND domain. However, in these data there is a unexplained excess at low energy over the (reliably?) estimated background.  In the neutrino data sample, only the events with neutrino energy above a threshold value $E_{th}$ were used for the search of a LSND-like signal (with negative result), leaving the issue of an explanation of the low energy excess unanswered. In the antineutrino channel most of the support to the LSND signal appears to arise from an excess above $E_{th}$ but quite close to it, so that there is, in my opinion, some room for perplexity. Then there is the reactor anomaly: a reevaluation of the reactor flux \cite{flux}, produced an apparent gap between the theoretical expectations and the data taken at small distances from the reactor ($\lappeq$ 100 m). The discrepancy is of the same order of the quoted systematic error whose estimate, detailed in the paper, should perhaps be reconsidered. Similarly the Gallium anomaly \cite{gal} depends on the assumed cross-section which could be questioned. The reactor anomaly and the Gallium anomaly do not really agree on the oscillation parameters that they point to: the $\Delta m^2$ values are compatible but the central values of $\sin^2{2\theta}$ differ by about an order of magnitude, with Gallium favouring the larger angle. The cosmological data appear compatible with at most the existence of 1 sterile neutrino (the most stringent bounds arising form nucleosynthesis). Over all, only a small leakage from active to sterile neutrinos is allowed by present neutrino oscillation data. If all the indications listed above were confirmed (it looks unlikely) then 1 sterile neutrino would not be enough and at least 2 would be needed with sub-eV masses. Establishing the existence of sterile neutrinos would be a great discovery. In fact a sterile neutrino is an exotic particle not predicted by the most popular models of new physics. A sterile neutrino is not a 4th generation neutrino (for the status of a 4th generation see the talk by Ivanov): the latter is coupled to the weak interactions (it is active) and heavier than half the Z mass. A sterile neutrino would probably be a remnant of some hidden sector. The issue is very important so that new and better experimental data are badly needed (MiniBooNE will present new results in the summer).

Some new physics hints from accelerator experiments were discussed at this Conference. Let me make a list of the most plausible ones. Since the LEP time we have the discrepancy of the forward-backward asymmetry of the b quark, $A^b_{FB}$, which is about 3$\sigma$ away from the SM fit. This discrepancy, if not of experimental origin, might indicate some anomaly in the $Z\rightarrow b \bar b$ vertex. Then we have the muon g-2 measurement. This is also at about the 3$\sigma$ level from the SM prediction, if we believe the theoretical error estimate (the main source of uncertainty is from the hadronic contribution to the light-by-light scattering contribution and, to a lesser extent, to the photon vacuum polarization). The g-2 measurement is so precise that it is realistic to imagine that a discrepancy in this measurement could be among the first detected signals of new physics. And in fact, for example, if we take the case of supersymmetry, we can easily reproduce this anomaly by relatively light values of EW s-partner masses and moderately large values of $\tan{\beta}$ without at the same time contradicting other available measurements \cite{ref:sus}. As already mentioned, 
the problem is that the LHC bounds on SUSY particles are now progressively pushing the favoured area in parameter space away \cite{struFT} (talks by Potter and Kiesenhofer, and, for the Tevatron searches, by Zivkovic). 

Another interesting hint arises from the Tevatron measurement of the forward-backward asymmetry of the t quark, $A^t_{FB}$, which, according to the CDF experiment, shows a discrepancy with the SM which is particularly prominent at large rapidity and large $t\bar t$ mass $M_{tt}$ (see the talks by Schwartz and Harel). At $M_{tt} \gappeq$ 450 GeV the CDF value is at about 3.5 $\sigma$ above the SM prediction. This high mass region is exciting because it is on the side where new physics could be present. But the D0 measurement at $M_{tt} \gappeq$ 450 GeV shows a much smaller excess. At the LHC one cannot measure the same quantity because, contrary to p$\bar p$, the pp initial state is forward-backward symmetric, but one can try to detect a different rapidity distribution for t and $\bar t$ quarks. But the predicted effect is small and, at present, the LHC experiments have not yet reached a sufficient sensitivity (see the talks by Tosi and Helsens).
From the theoretical point of view the discussion is progressing on two different fronts. One has to do with the question whether the available theoretical prediction of the effect in the SM is really solid. The second is on the possible new physics mechanisms that could explain the observed asymmetry. The first question was discussed by Vogelsang in his talk. One might worry about the QCD prediction of the asymmetry because the asymmetry numerator is only computed at the leading non vanishing order. Moreover the NLO correction to the denominator (i.e. the $t \bar t$ production cross section) leads to a $ \sim 30 \%$ effect due to a rather large scale uncertainty. However, the inclusion of the resummed  leading log-enhanced higher order contributions affects numerator and denominator in about the same way, so that Vogelsang's conclusion was that, if the data persist, it is unlikely that the QCD higher order terms can explain the discrepancy. The possible new physics explanations were discussed by Kamenik. An exotic particle exchanged in the s-channel must be a colour octet (to interfere with gluon exchange) with non vanishing axial coupling. One has considered axi-gluons (but also scalar octets). In addition the $u \bar u$ and $t \bar t$ couplings must be of opposite sign to lead to a positive asymmetry.  These models are increasingly constrained by LHC searches of high mass di-jet events.
Alternatively one can consider a new Z' or W' exchanged in the t-channel, with flavour changing couplings that turn a valence (anti)quark of the (anti)proton into a (anti)top quark. The predicted $M_{tt}$ distribution is normally slowly rising. Large flavour changing couplings are needed. Then in addition to the desired $u \bar u \rightarrow t \bar t$ amplitude one also would get the dangerous $u u \rightarrow t  t$ amplitude leading to double top production and thus, in a fraction of cases, to equal charge di-leptons. Experiments pose a strong constraint on models of this type. So the issue remains open. 

Another possible indication of new physics comes from the like sign di-muon asymmetry measured by D0 and presented here by Williams. Interpreted as arising from CP violation in the $B_{d,s}$ semileptonic decays it amounts to a 3.2 $\sigma$ discrepancy. This measurement cannot be performed by CDF because the magnetic field cannot be inverted. The issue is delicate because the $p \bar p$ initial state is CP symmetric but the detector is made by matter and is not CP symmetric. In his talk Lenz has argued that the central value of the asymmetry measured by D0 exceeds the theoretical bound obtained in the assumption that the non diagonal term $\Gamma_{12}$ is non exotic, which is sensible because only light (few GeV) intermediate states can contribute to the imaginary part of the box diagram. It is interesting that relatively soon a different combination of the relevant semileptonic asymmetries can be measured by LHCb, as discussed by Calvi, and the result of D0 will be checked.

The only direct indication of the production of a new particle is the CDF signal of a bump at $\sim145$ GeV in the di-jet mass distribution observed in Wjj final states (talk by Cuelca Almenar). One may wonder how is it possible that such a low mass state has not already been detected elsewhere. It is clear that this can only be obtained at the price of accepting a peculiar set of couplings for this particle. D0 has searched for this signal and has not confirmed it. Some skepticism is in order.

In B decays much attention is focussed on the study of $B_s \rightarrow J/\Psi \phi$ (talks by Abbot, Calvi, Muheim, Ruffini and Ben Haim). Initially CDF and D0 both reported a significant departure from the SM prediction. More recently the discrepancy has decreased in the new data: the last figures (at the EPS Conference in Grenoble in July '11) show only a deviation by about 1$\sigma$, in both CDF and D0, separately and in the same direction. Recently LHCb has released its first results, based on 37$fb^{-1}$ on the same decay. Also in this case a hint of a deviation by a little more than  1$\sigma$ is observed. LHCb will soon increase its statistics by more than a factor of 10 and so the issue will be clarified. $B_s \rightarrow \mu \mu$ is another channel under experimental study, with a sensitivity which is coming closer to the SM value, where a signal of new physics could appear (talk by Mancinelli). 

In the $B_d$ sector some tensions are also reported, for example, between the value of $\sin{2\beta}$ measured in $B_d$ mixing and the result of the CKM fit, as discussed by Lenz in his talk. A related discrepancy that enters in determining the goodness of the overall CKM fit is the deviation of the $V_{ub}$ values obtained from inclusive and exclusive channels: $V_{ub}^{incl} = 4.35\pm0.18(exp)\pm0.23(th)$ to be compared with  $V_{ub}^{excl} = 3.25\pm0.12(exp)\pm0.28(th)$, which amounts to about 2.6$\sigma$ (talk by Bernlocher). Note that the theoretical errors are dominant. Their determination is not really a rigorous matter. I think that this tension may be due to the fact that over the last 30 years hundreds of theory papers have been devoted to the determination of $V_{ub}$ with each author claiming that his/her work led to a decrease of the theoretical error, finally resulting in an underestimate of the stated ambiguity.

As a final example of possible deviations from the SM predictions we can mention the case of the $B \rightarrow \tau \nu$ decay (talk by De Nardo). The combined value from BaBar and Belle is about 2$\sigma$ away from the SM prediction. The exchange of a charged Higgs could produce such an effect.

The experiments on lepton flavour violation, like the experiment MEG that is looking for $\mu \rightarrow e \gamma$, are extremely important (talks by Miyazaki and De Gerone). MEG is about to release a new limit on this process (in fact, the limit on the branching ratio went from the previous value of 1.1 $10^{-11}$ down to the new figure of 2.4 $10^{-12}$, presented at the EPS Conference in Grenoble in July '11).

In summary, there a number of indications for new physics in accelerator experiments. However, it seems to me that none of those hints, for one reason or another, is really compelling. Once neutrino masses have been included in the main framework by a limited enlargement of the minimal SM, the most solid evidence for new physics is from cosmology: Dark Matter, Dark Energy, baryogenesis. But in these cases the relation with the EW scale is unclear. Dark Matter could be due to axions, baryogenesis could occur through leptogenesis (from the decay of heavy Majorana neutrinos with the intermediary of instanton effects) and the Dark Energy solution
could come from a far away layer of physics. It remains for the LHC and the present and future flavour factories (talks by Ratcliff, Bowder, Collins, Nomura and Casey) to widen the scope and the sensitivity of the search of new physics near the EW scale.

\section{Conclusion}

This was a very interesting Conference with a limited number of participants, all experts in the domain of flavour physics that plays a crucial role in particle physics today. There was large space for discussions, also outside the lecture room, and the setting was very pleasant. The hope is that for next edition of FPCP the LHC will have clarified some of the issues which are open today.

\Acknowledgements

I would like to thank the Organizers of FPCP 2011, in particular Abner Soffer, for the kind invitation and warm hospitality.
I recognize that this work has been partly supported by the Italian Ministero dell'Universit\`a e della Ricerca Scientifica, under the COFIN program (PRIN 2008).

\end{document}